\documentclass[9pt,twocolumn,twoside]{opticajnl}
\journal{opticajournal} 

\setboolean{shortarticle}{true}

\usepackage{amsmath,amssymb,amsfonts,commath}
\usepackage{lineno}
\usepackage{gensymb}

\title{On-chip GaAs carrier lifetime investigation}

\author[1*,2]{Chawaphon Prayoonyong}
\author[2]{Crispin Szydzik}
\author[2]{Guanghui Ren}
\author[1,2]{Bill Corcoran}
\author[2]{Aditya Dubey}
\author[1,2]{Caitlin E. Murray}
\author[2]{Toby Mitchell}
\author[2]{Arnan Mitchell}

\affil[1]{Photonic Communications Laboratory, Dept. Electrical and Computer Systems Engineering, Monash
University, Clayton, VIC, Australia}
\affil[2]{Integrated Photonics and Applications Centre (InPAC), Melbourne, VIC, Australia}

\affil[*]{park.prayoonyong@monash.edu}

\begin{abstract}
We propose an alternative way to determine GaAs carrier lifetime using pump-probe measurement based on fibre optics and integrated waveguides. We find that our GaAs samples have the lifetime ranging from 30-80 ps, supporting the bandwidth $\geq$ 12.5 GHz. The platform utilised in this work could offer a cost-effective way to investigate photocarrier lifetime. Moreover, it may have the potential for high-speed switches and detectors, or could be exploited for possible all-optical modulation.
\end{abstract}

\setboolean{displaycopyright}{false} 

\begin{document}

\maketitle

\section{Introduction}
GaAs is semiconductor material used widely for applications in optoelectronics and photovoltaics owing to the direct bandgap structure with the gap energy corresponding to the wavelength of $\sim$800 nm and high charge mobilities \cite{yoon2010gaas,li2022performance}. These properties enable the compound and its alloys to be exploited in e.g. high-speed detectors \cite{chen2018integration,li2022high}, photovoltaic cells \cite{yoon2010gaas}, and terahertz (THz) applications \cite{lepeshov2017enhancement,gorodetsky2021operation} where terahertz waves are generated by semiconductor samples being illuminated with a short-pulse laser when biased with a DC voltage.

When optimising the performance of GaAs samples for any application, one should know carrier lifetime of the samples which is defined as the time before free carriers recombine after excitation. The quantity can be used to design the frequency response and bandwidth of optoelectronic devices. For photoconductive applications, one prefers the short lifetime in the order of picoseconds or below so that electrical output or emitted THz signals can follow the input laser pulses \cite{lepeshov2017enhancement,gorodetsky2021operation,chen2018integration,li2022high}. Meanwhile, carrier lifetime should be adequately long to allow carriers transport between the GaAs layer and metallic contacts in photovoltaic cells \cite{yoon2010gaas}.   

When determining the carrier lifetime of GaAs, one could deploy techniques of time-resolved reflectance or transmittance where a pulse whose energy is above the bandgap energy (pump) is used to excite carriers giving rise to changes in the refractive index that minimally alters how a low-power beam (probe) reflects from, or propagates through the samples, i.e. pump-probe measurements \cite{fischer2016invited}. To minimise unwanted interference between the pump and probe signals, the two signals are set to different polarisations with the pump at one polarisation having significantly higher power \cite{jani2020time,sabbah2002femtosecond,mcintosh1997investigation}, different optical wavelengths \cite{antoncecchi2020high,wolfson2018long,zhang2018ultrafast,di2020broadband} or optical and THz pulses \cite{tielrooij2013photoexcitation,zou2020carrier,mag2016low}. To the best of our knowledge, these pump-probe measurements are based on free-space optics where each device is expensive and should be carefully aligned to enhance the measurement results.

In contrast to previous experiments where conventional free space optics is deployed, in this paper, we propose an approach using fibre optics and integrated grating couplers to perform pump-probe measurements for carrier lifetime characterisation of GaAs samples. Since fibre-based devices are employed, the approach offers a more cost-effective and convenient way to investigate the rise and fall time for photocarriers. Moreover, the method requires low power for sample excitation due to high power concentration inside a fibre core. The results imply that our GaAs samples and integrated waveguides may have the potential for photoconductive switches or detectors with the bandwidth $\geq$ 12.5 GHz. Alternatively, the platform used in this paper could offer opportunities for all-optical modulation.


\section{Proof of concept}
\begin{figure*}[htbp]
    \centering
    \includegraphics[width=0.95\textwidth]{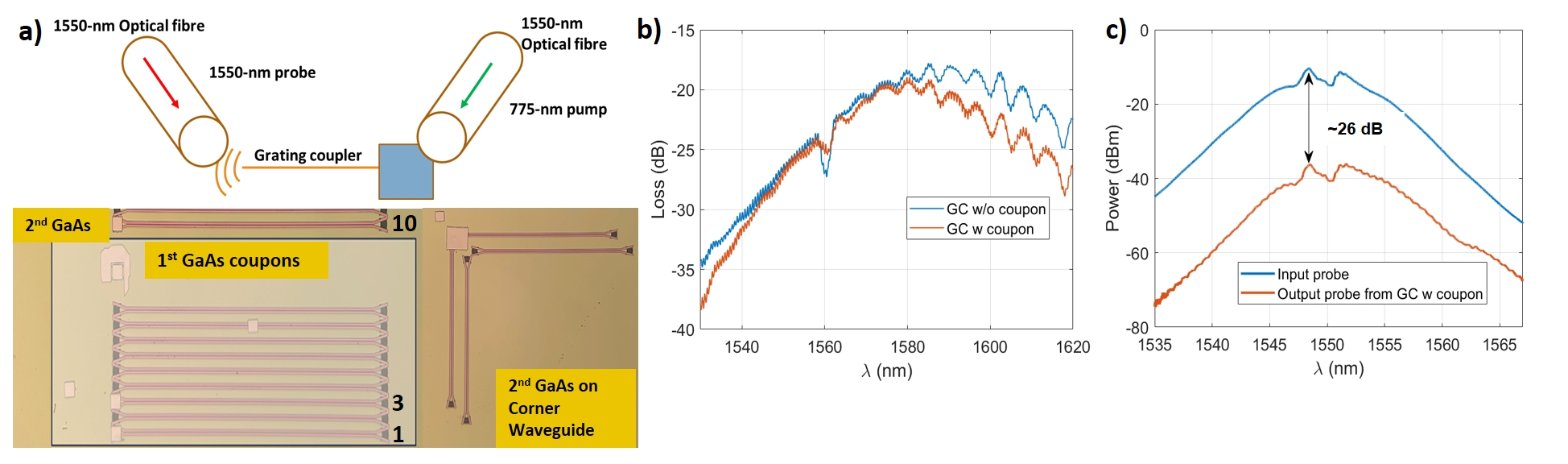}
    \caption{(a) (top) Schematic showing how we focus the pump and guide the probe, (bottom) Layout for GaAs coupons and grating coupler waveguides with GaAs coupons from the first batch are printed on GCs of waveguide 1 and 3, while those from the second batch are on top of GC waveguides 10 and at the chip corner, (b) Loss profiles of the waveguides against wavelength with and without GaAs coupons, (c) Spectra of the input pulse before and after being transmitted through the waveguides with GaAs coupons.}
    \justifying
    \label{fig:layout}
\end{figure*}

Fig. \ref{fig:layout}(a) shows how we focused the pump and guided the probe into a GaAs coupon. Here, four GaAs coupons (30$\mu m$ $\times$ 30$\mu m$) from two batches (1st and 2nd) were printed on top of grating couplers connected with SiN integrated waveguides \cite{han2022integrated}. Since the waveguide was designed for wavelengths $\sim$1550 nm, we used it to guide the lower-energy probe with the aid of the other grating coupler. On the coupon, we illuminated the pump pulse at wavelengths $\sim$775 nm to excite electrons into the conduction band and, in turn, change the coupon transmittance. This lasts until the charges relax back to the valence band. 

 For wavelength dependence, it is seen that the grating couplers (GC) with coupons on top have the similar response to those without as shown in Fig. \ref{fig:layout}(b). When launching a pulse at 1550 nm through the platform (GC+coupon), the output spectrum seems negligibly distorted (Fig. \ref{fig:layout}(c)). We see that output spectrum experiences similar loss value to that at 1550 nm in Fig. \ref{fig:layout}(b) $\sim$26 dB. 

\begin{figure}
    \centering
    \includegraphics[width=0.75\columnwidth]{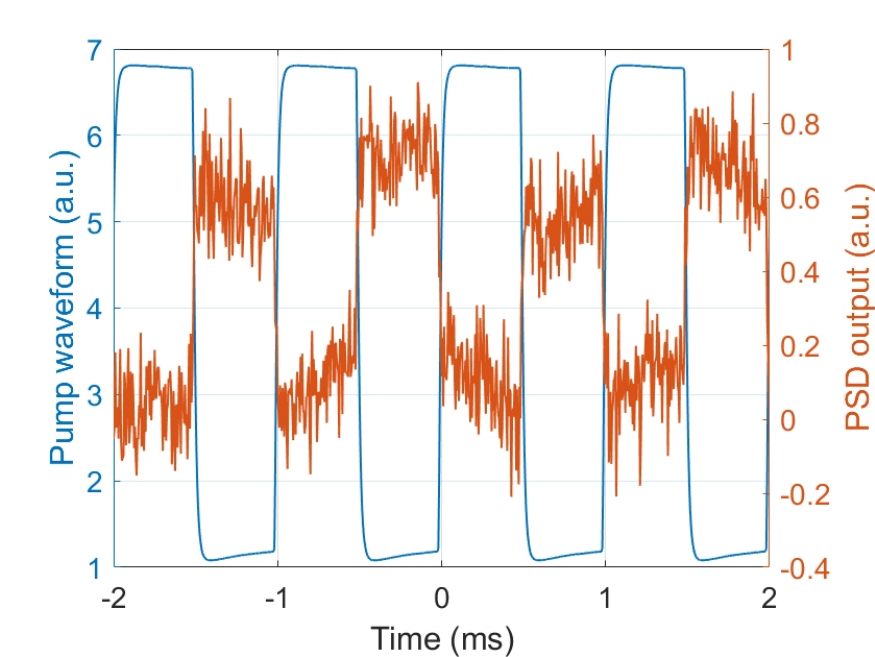}
    \caption{Slowly modulated 780-nm laser as the reference signal (blue), and the received 1550-nm laser after passing through the coupon (red).}
    \justifying
    \label{fig:CW_test}
\end{figure}

To verify the platform's feasibility, we shined a 780-nm laser modulated with a square wave at a rate of 1 kHz on the GaAs coupon, while launching a 1550-nm CW laser through the grating coupler. By using an oscilloscope and slow detector, we can see the 780-nm and received 1550-nm signals as seen in Fig. \ref{fig:CW_test}. Here, the received signal follows the reference with 180$^{\circ}$ phase shift, indicating that the transmittance of the sample changes when being illuminated with the 780-nm light. This implies that the platform could be deployed for the lifetime investigation. When both signals are pulse lasers, we expect to see the largest change in transmission of the 1550 nm probe when the delay between pump and probe is zero.  
When performing pump-probe measurement on the samples, the number of photoinduced carriers after excitation by the pump pulse ($\Delta n$) could be expressed as in Eq. \ref{eq:recomb} 
\begin{equation} \label{eq:recomb}
    \frac{d}{dt}\Delta n=-\frac{\Delta n}{\tau_{1/e}}+G
\end{equation}

Here, $\tau_{1/e}$ is the carrier lifetime before recombination and $G$ is the generation rate due to the number of incoming photons which is related to the pump intensity profile. By solving Eq. \ref{eq:recomb}, the number of photoinduced carriers $\Delta n$ is the convolution of pump intensity, $G$, and carrier decay function ($\propto exp(\frac{t}{\tau_{1/e}})$). Ideally, if the pump pulse were the Dirac delta function ($\delta(t)$), we would see a perfect one-sided exponential decay of $\Delta n$ and could calculate for carrier lifetime directly. To achieve this, the pump pulse should be much shorter than the carrier lifetime. At the same time, the probe pulse should also be very short to ensure that the convolution response of transmitted probe against delay time is close to an exponential decay \cite{fischer2016invited,dong2017pump,sabbah2002femtosecond}.
\section{Experimental setup}

\begin{figure}[htbp]
    \centering
    \includegraphics[width=\columnwidth]{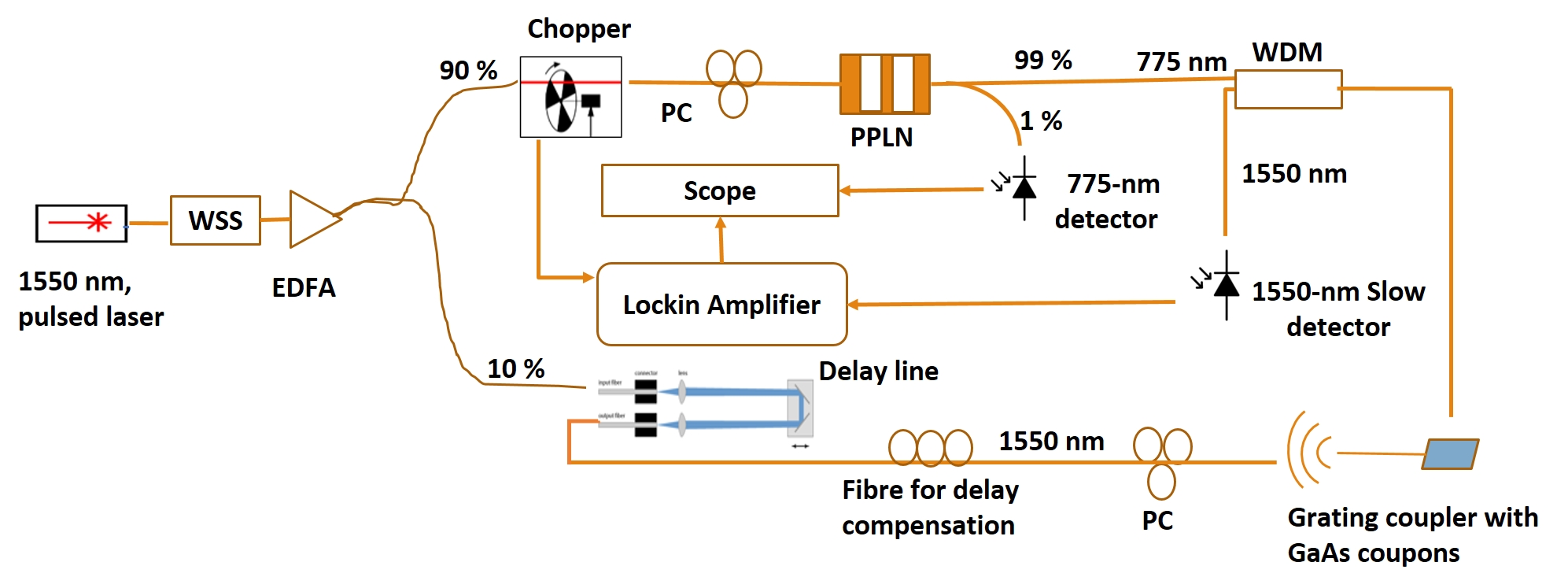}
    \caption{Experimental setup. Here, WSS:Wavelength Selective Switch (optical filter), EDFA:Erbium-dope fibre amplifier, PC:Polarisation Controller, PPLN:Periodically-Poled Lithium Niobate, WDM:Wavelength Division Multiplexer.}
    \label{fig:Exp_setup}
\end{figure}

Fig. \ref{fig:Exp_setup} illustrates the experimental setup employed to determine the carrier lifetime. Here, the pulsed laser (Pritel fibre laser) centre around the wavelength of 1552 nm with a repetition rate of 40 MHz was launched into an optical filter (WSS - \textit{Finisar} Waveshaper) and optical amplifier (EDFA) to vary the pulse width and boost the average power to 13 dBm, respectively. The output was then split into 90\% and 10\% parts using a coupler. The 90\% of power propagated through collimators and an optical chopper operating at 730 Hz, leading to a slow square signal used as the reference for a lock-in amplifier (Ametek Inc.). The signal was fed into a second harmonic generator based on periodically-poled lithium niobate (PPLN) from AdvR Inc. to convert pulses at the wavelength of 1552 nm to 776 nm. Since the PPLN device was polarisation selective, a polarisation controller (PC) was employed to align the input polarisation to the device, and to maximise the output power. 
The 776-nm pulse was transmitted to illuminate a GaAs coupon via a 775/1550-nm wavelength division multiplexer (WDM).
\begin{figure*}[htbp]
    \centering
    \includegraphics[width=\textwidth]{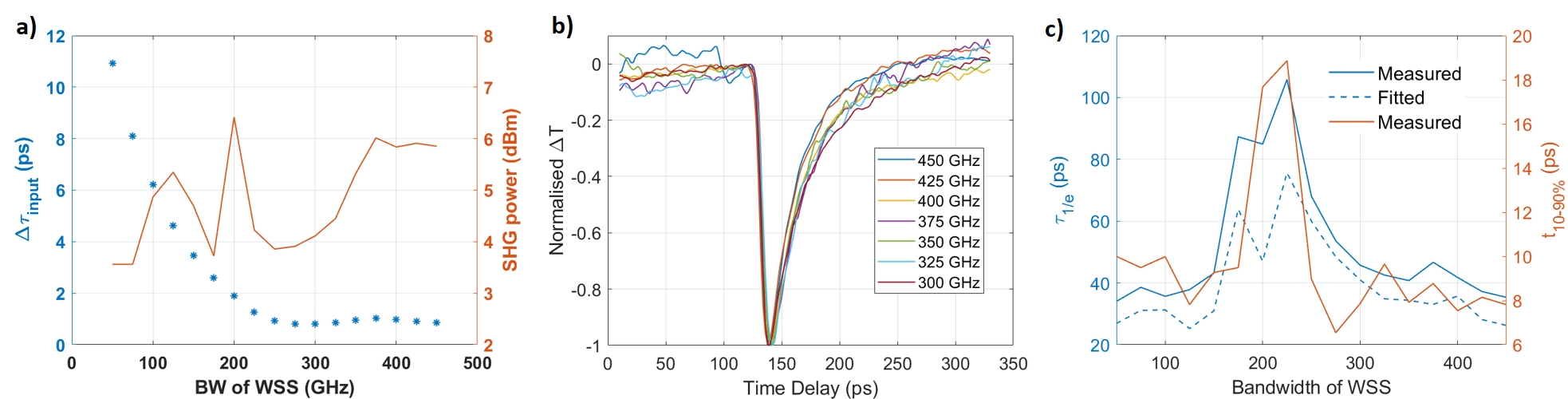}
    \caption{(a) Autocorrelation widths of the input pulse and the corresponding second harmonic (SH) power against the WSS bandwidth, (b) Normalised transmittance ($\Delta T$) of the GaAs coupon on waveguide no. 1 when adjusting the delay line and WSS bandwidth, (c) Lifetime measured and fitted from the transmittance traces along with the measured 10-90\% rise time.}
    \label{fig:BW_and_results}
\end{figure*}

Meanwhile, the 10\% of pulse power exploited as the probe proceeded to two motorised delay lines to adjust the delay with a step of 2 ps between the pump and probe. We note that extra optical fibre was also deployed to ensure that the delay between the pulses was within the delay lines range (660 ps). The probe signal was then sent to the grating coupler with a PC aligning input polarisation to that of the waveguide. Before the pump was launched, we aligned bare fibre ends with grating couplers to minimise loss, helping align the pump and probe.

After the probe signal was modulated due to the changes in transmittance caused by the pump, it was received at a slow detector (Thorlab) for 1550-nm wavelengths via the 775/1550-nm WDM. The received signal was amplified and multiplied with the reference signal inside the lockin amplifier which could extract low-power signals from noisy conditions and return quasi-DC output. We expected to see the largest change in DC output when the pump and probe were aligned. During the experiment, an oscilloscope was used to monitor and compare signals from the chopper and amplified input of the lock-in amplifier. In addition, the reference phase of the lock-in amplifier was set so that received signals with the opposite sign to the reference (e.g. Fig.\ref{fig:CW_test}) manifest negative DC at the output. 


\section{Results}
To ensure that we obtained optimal pulse width for this experiment, we varied the bandwidth of WSS, while fixing the applied dispersion, and measured the autocorrelation width using an autocorrelator. Meanwhile, the power of SH signal after the pulse was fed to PPLN was monitored with the 775-nm photodetector. Results for the autocorrelation width and corresponding SH power are shown in Fig. \ref{fig:BW_and_results}(a). Intuitively, the autocorrelation width becomes smaller when the WSS bandwidth increases. However, it flattens at around 1 ps when the bandwidth is more than 250 GHz. We suspect that amplification and nonlinearity, e.g. self-phase modulation, inside the EDFA may introduce unwanted chirps preventing the pulse from being shorter than this limit. At the same time, the corresponding SH power also varies with the bandwidth ranging from $\sim$ 4-6 dBm, considered to be 12.5-20 \% for the conversion efficiency (CE) of SHG at 13 dBm of the pulsed input. This is in contrast to the CE of 1 \% when the input CW laser at 16 dBm is launched to the PPLN device, according to the specification. 

Since our autocorrelator did not operate at 775-nm wavelengths, we could not optimise the SH pulse width in the experiment using the device. Instead, we ran the pump-probe experiment on the GaAs coupon no.1, while sweeping the WSS bandwidth for the input pulse to optimise the measurement results of transmittance profiles against time delay. Fig. \ref{fig:BW_and_results}(b) shows the normalised transmittance profiles against time delay when the WSS bandwidth was swept. When the pump and probe were aligned the lock-amplifier output became more negative in this case. Before normalisation, we note that there was the nonzero background before the pulses were aligned, and after the decay saturated. This might be cumulative heat caused by the pump, resulting in a refraction index change \cite{antoncecchi2020high}. As seen, the profiles seem to follow an exponential decay as discussed in Eq.\ref{eq:recomb}. We then calculate the 10-90\% rise time and lifetime, $\tau_{1/e}$, of the traces.

Fig. \ref{fig:BW_and_results}(c) shows the lifetime measured from the width of a transmittance profile at the 37\% ($1/e$) and exponential fitting, when varying the WSS bandwidth. We see that the lifetime read from measurement seems to saturate around 40 ps when the bandwidth is set in ranges of 50-150 GHz and 300-450 GHz. The result from fitting exhibits a similar trend but with $\sim$ 10 ps lower in these ranges as we ignore the rising edges in the calculation. The measured 10-90\% rise time in these ranges lies between $\sim$ 6-10 ps, which may be due to the pulse resolution. When looking at the result in a range of 175-275 GHz, the lifetime is significantly large exceeding 100 ps read from measurement and 60 ps from fitting. In addition, the 10-90\% rise time in this range also goes up to almost 20 ps. This implies that the SH pump pulse may have a broad profile. With the optimal WSS bandwidth and exponential fitting, it is inferred that carrier lifetime should be $\sim$30 ps for Coupon 1.    
\begin{figure}[htbp]
    \centering
    \includegraphics[width=0.78\columnwidth]{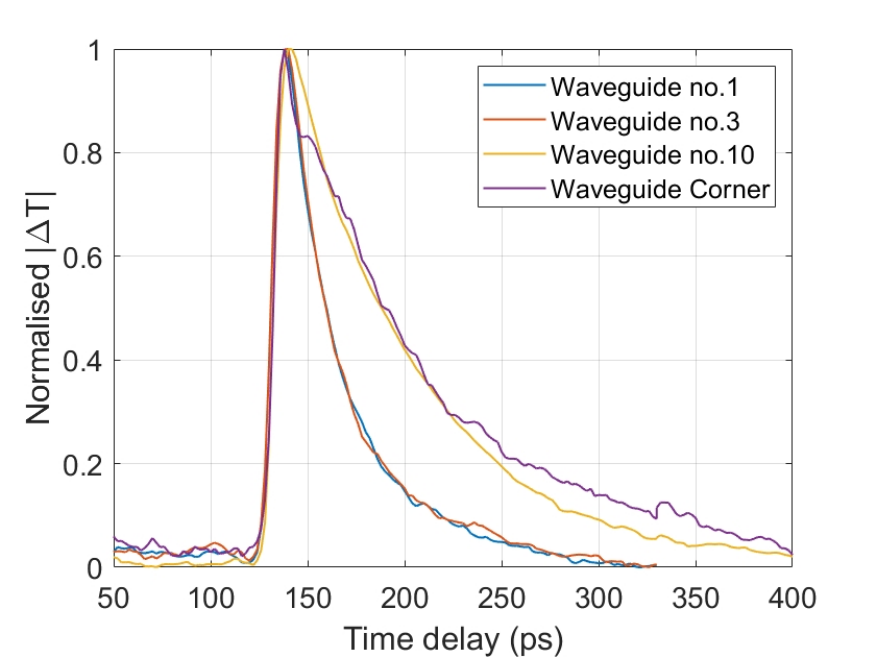}
    \caption{Normalised magnitude of transmittance changes against time delay between pump and probe signals for different GaAs coupons}
    \label{fig:alltime}
\end{figure}

We then performed the pump-probe experiments on all the coupons when setting the WSS bandwidth to 425 GHz. The magnitude of transmittance changes for all GaAs coupons against time delay is shown in Fig. \ref{fig:alltime}. We note that only Coupon 10 exhibits a positive change in transmittance, while the others manifest negative changes when the pulses are aligned in time. We think that Coupon 10 might experience stimulated emission while others see excited stimulated absorption. This might be because Coupon 10 was printed on the GC in a way that the pump and probe signals were aligned differently from others. The deeper investigation of this phenomenon is out of our scope.

As seen in Fig. \ref{fig:alltime}, the trace from Coupon 1 is similar to that from Coupon 3, while the trace of Coupon 10 is similar to that of the Corner coupon. This might be because they were taken from the same batches. We summarise the coupon transmittance in Table. \ref{tab:time}.
\begin{table}[htbp]
\centering
\caption{\bf Transmittance lifetime and 10-90\% rise time}
\begin{tabular}{ccc}
\hline
 Coupon & fitted $\tau_{1/e}$ (ps) & $t_{10-90\%}$ (ps) \\
\hline
1& 31.62& 7.23 \\
3& 31.17& 8.28 \\
10& 66.34& 8.29 \\
Corner& 76.81& 7.62 \\
\hline
\end{tabular}
  \label{tab:time}
\end{table}

From the table, GaAs samples from the first batch (Coupon 1 \& 3) manifest a carrier lifetime of 30-40 ps, while those from the second batch (Coupon 10 \& Corner) exhibit a lifetime of 70-80 ps. The lifetime differences between samples from the same batches may be as a result of noise and the fitting process. The results are similar to those acquired from the optical pump-THz probe technique \cite{zou2020carrier}. For the 10-90\% rise time, all coupons manifest a similar number around 6-8 ps, which may be limited by the pump resolution. 



\section{Conclusion and Discussion}

By using integrated grating coupler and fibre optics devices, we can achieve pump-probe measurements for GaAs samples with the carrier lifetime below 100 ps. This may be a result of defects or unintentional doping causing the period to be shorter than that of intrinsic GaAs ($\geq 1$ ns).

From the results, our GaAs samples could be intrinsically deployed for photoconductive switches with sub-THz frequency as the carriers can only follow a laser pulse with a width $\geq$ 30 or 80 ps (i.e. 12.5-33 GHz bandwidth). To operate at higher frequency, one could introduce structures like nanoantennas or finger electrodes to reduce the lifetime on top of GaAs substrate \cite{lepeshov2017enhancement,chen2018integration}. 

In addition, integrated grating couplers with GaAs coupons printed on top could have potential for all-optical modulation where 1550-nm signals could be intensity modulated by 775-nm signals via a GaAs coupon, but the modulation strength might be minimal. If the grating coupler waveguide is designed for 775-nm wavelengths and GaAs coupon is connected with electrodes for DC bias, one could employ the platform as photoconductive switches or detector \cite{chen2018integration,li2022high}. To enable THz sampling or emitting, one could exploited Low-temperature-grown GaAs for the coupon, or introduce finger structures on the substrate.

This approach offers a more cost-effective and less-effort way than the conventional methods where free-space optics devices are more expensive and difficult to align or ensure that the spot size and pump power are sufficient for photoexcitation \cite{antoncecchi2020high,jani2020time}. This is in contrast to our approach where one can connect fibre-based devices easily and focus the pump light into samples with the diameter of fibre core ($\leq10 \mu m$). In our case, the power concentration on the coupons could be 5.1 kW/cm$^2$ based on the average pump power of 6 dBm at 775 nm. However, the approach comes with drawbacks one should consider since integrated waveguides are required to align the signals, and fibre dispersion may spread the pulses.

\begin{backmatter}
\bmsection{Funding} Awaiting details of the funding information.

\bmsection{Acknowledgments} 
CP and AM would like to thank Prof. Yasuhiro Tachibana from School of Engineering, RMIT University for Spectroscopy laboratory tour, and fruitful discussion. Device fabrication was carried out within the RMIT Micro Nano Research Facility (MNRF) in the Victorian Node of the Australian National Fabrication Facility (ANFF-Vic).

\bmsection{Disclosures} 
The authors declare no conflicts of interest.

\smallskip











\bmsection{Data availability} Data underlying the results presented in this paper are not publicly available at this time but may be obtained from the authors upon reasonable request.






\end{backmatter}

\bibliography{sample}

\bibliographyfullrefs{sample}


\ifthenelse{\equal{\journalref}{aop}}{%
\section*{Author Biographies}
\begingroup
\setlength\intextsep{0pt}
\begin{minipage}[t][6.3cm][t]{1.0\textwidth} 
  \begin{wrapfigure}{L}{0.25\textwidth}
    \includegraphics[width=0.25\textwidth]{john_smith.eps}
  \end{wrapfigure}
  \noindent
  {\bfseries John Smith} received his BSc (Mathematics) in 2000 from The University of Maryland. His research interests include lasers and optics.
\end{minipage}
\begin{minipage}{1.0\textwidth}
  \begin{wrapfigure}{L}{0.25\textwidth}
    \includegraphics[width=0.25\textwidth]{alice_smith.eps}
  \end{wrapfigure}
  \noindent
  {\bfseries Alice Smith} also received her BSc (Mathematics) in 2000 from The University of Maryland. Her research interests also include lasers and optics.
\end{minipage}
\endgroup
}{}

\end{document}